 \documentclass[final,5p,times]{elsarticle}

\usepackage{amssymb}

\usepackage{lineno}

\usepackage{listings}

\journal{SoftwareX}

\begin{document}

\begin{frontmatter}



\title{TBTK: A quantum mechanics software development kit}


\author{Kristofer Bj\"{o}rnson}
\address{Niels Bohr Institute, University of Copenhagen, Juliane Maries Vej 30, DK-2100 Copenhagen, Denmark}

\begin{abstract}
TBTK is a software development kit for quantum mechanical calculations and is designed to enable the development of applications that investigate problems formulated on second-quantized form.
It also enables method developers to create solvers for tight-binding, DFT, DMFT, quantum transport, etc., that can be easily integrated with each other.
Both through the development of completely new solvers, as well as front and back ends to already well established packages.
TBTK provides data structures tailored for second-quantization that will encourage reusability and enable scalability for quantum mechanical calculations.
\end{abstract}


\end{frontmatter}


\section{Introduction}
\label{Section:Introduction}
For more than half a century technological progress has been fueled by advances in semiconductor technology, with exponential progress described by Moore's law.
The main driving factor behind this is the continuous decrease in transistor size.
The International Technology Roadmap for Semiconductors (ITRS) targets 5 nm technology in 2021 (http://www.itrs2.net/), but further decrease is difficult.
With a lattice constant of 0.54 nm, a silicon cube with side length 5 nm is roughly nine unit cells wide and contains no more than a few thousand silicon atoms.
On this scale quantum mechanical effects starts to dominate~\cite{ElectronicTransportInMesoscopicSystems, AtomToTransistor}.
Easy access to more accurate models are therefore needed to complement the semi-classical models that have been sufficient for industrial purposes in the past~\cite{PhysicsOfSemiconductorDevices}.

Simultaneously, increased computational power, in combination with algorithm development, has increased the number of atoms that can be simulated using quantum mechanical models \cite{ALPSProject, Wien2k, Gaussian, Vasp, PhysRevLett.105.167006}.
The system sizes that are accessible for both academic and industrial interests are therefore already overlapping.
This is not least visible through recent advances in the field of quantum computing, where academia and industry are making significant advances together \cite{Nature.541.447, Nature.553.136, Nature.543.159, Nature.549.149}, as well as through the increased governmental spending on quantum technologies~\cite{QuantumFlagship, NationalQuantumInitiative}.
For this effort to be successful, it is important to develop tools and procedures that enables subject experts from different fields to collaborate effectively with each other.
In particular, data structures that provide high level abstractions of quantum mechanical quantities are needed.

Consider the notation $|\Psi_{lm\sigma}(x, y, z)\rangle$, which is mathematically eqivalent to a representation using the notation $|\Psi_{h}\rangle$, where $h$ is some linear Hilbert space index.
The former representation is a high level abstraction particularly suited for model specific reasoning, while the later is a low level representation suited for method developers interested in implementing computationally demanding general purpose algorithms.
Without a general mapping between the two representations, low level design decissions are bound to propagate upward in the code.
They can either propagate all the way to the end user or be hidden away through a high level, application specific, interface.
In the former case, responsibility is put on the end user to understand the low level conventions, while in the later case the generallity of the code likely is limited.
In either case, the code is made difficult to integrate with other softwares since a set of universally agreed upon conventions are lacking.

In this paper we present TBTK, an SDK for modeling and solving Hamiltonians on second-quantized form.
At its core it solves the mapping problem described above.
It also provides an extensive set of general purpose data structures that can be used to implement new applications, solvers, as well as front ends and back ends to already existing packages.

\section{Data structures: abstraction and efficiency}
The main part of TBTK is a C++ library that contains data structures meant to simplify both the development of applications that investigate particular quantum mechanical questions, as well as enable developers to implement general purpose reusable solvers.
The data structures are designed to provide abstractions that allow the developer to focus on physics instead of numerics, and to provide the same efficiency as highly optimized single purpose codes.
Through strong emphasis on object oriented design, the code is divided into logical units protected by strong encapsulation, which enables developers to work on the level of abstraction appropriate for the given task.

\section{Second quantization}
The starting point for TBTK applications are Hamiltonians in second quantized form
\begin{equation}
	H = H_0 + H_{I} = \sum_{\mathbf{i}\mathbf{j}}a_{\mathbf{i}\mathbf{j}}c_{\mathbf{i}}^{\dagger}c_{\mathbf{j}} + H_{I},
\end{equation}
where $a_{\mathbf{i}\mathbf{j}}$ are complex numbers, $\mathbf{i}$ and $\mathbf{j}$ are discrete indices and $c_{\mathbf{i}}^{\dagger}$ and $c_{\mathbf{i}}$ are creation and annihilation operators, respectively, for state $\mathbf{i}$.
Preliminary support is available for interaction terms $H_{I}$, but in this brief introduction we focus on non-interacting Hamiltonians.
In TBTK notation, the complex numbers $a_{\mathbf{i}\mathbf{j}}$ are called hopping amplitudes, which is derived from the codes initial focus on tight-binding calculations.
However, the nomenclature is more generally motivated by the fact that when the Schr\"{o}dinger equation
\begin{equation}
	i\hbar\partial_{t}|\Psi(t)\rangle = H_{0}|\Psi(t)\rangle
\end{equation}
is rewritten using finite differences
\begin{equation}
	|\Psi(t+dt)\rangle = (1 - \frac{idt}{\hbar}\sum_{\mathbf{i}\mathbf{j}}a_{\mathbf{i}\mathbf{j}}c_{\mathbf{i}}^{\dagger}c_{\mathbf{j}})|\Psi(t)\rangle,
\end{equation}
the $a_{\mathbf{i}\mathbf{j}}$'s are seen to be amplitudes associated with the process whereby particles are annihilated in state $\mathbf{j}$ and recreated in state $\mathbf{i}$.
That is, the particle is hopping from state $\mathbf{j}$ to state $\mathbf{i}$.

\section{Physical indices and Hilbert space indices}
To allow for Hamiltonians of arbitrary complexity to be specified, TBTK provides a flexible indexing scheme.
An important distinction is made between physical indices such as (x, y, z, sublattice, orbital, spin) that have an intuitive connection to the physical problem at hand, and Hilbert space indices that are linear representations of the corresponding physical indices.
A typical mapping from a physical index to a Hilbert space index, say for a two dimensional lattice with a spin having the index structure (x, y, s), could be hard coded into an application as h = 2*SIZE\_Y*x + 2*y + s.
The problem with such an explicit mapping is that it forces every aspect of the application to work with this convention, both limiting the applicability of the code and leaving unnecessary numerical details visible at every level of the code.
TBTK solves this through a combination of flexible indices and a sophisticated storage structure for the hopping amplitudes and indices that automatically provides an efficient mapping between the physical indices and the Hilbert space indices.
Application developers can therefore work with physical indices exclusively, while method developers can write general purpose solvers that only depends on the Hilbert space indices.
In TBTK a physical index is specified using curly braces such as \{x, y, s\}.

For full generality, TBTK also allows for indices with different index structures to be used simultaneously, as for example is the case for a system that consists of two subsystems with index structure (x, s) and (x, y, s), respectively.
For this to be possible, the only requirement is that the indices differ in a subindex to the left of where the index structure first differs.
This is easily solved by adding a subsystem index at the front, resulting in the numerical indices \{0, x, s\} and \{1, x, y, s\}.

\section{Creating models}
A hopping amplitude is uniquely determined by its complex value and the two indices $\mathbf{i}$ and $\mathbf{j}$.
A hopping amplitude with value 1 from state \{x, y, s\} to \{x+1, y, s\} is created using
\begin{lstlisting}[language=c++]
HoppingAmplitude(1, {x+1, y, s}, {x,y,s});
\end{lstlisting}
Further, all hopping amplitudes are stored inside an object called Model and for example a square lattice with nearest neighbor hoppings can be set up as follows.
\begin{lstlisting}[language=c++]
Model model;
for(int x = 0; x < SIZE_X; x++){
  for(int y = 0; y < SIZE_X; y++){
    for(int s = 0; s < 2; s++){
      if(x+1 < SIZE_X){
        model << HoppingAmplitude(
          1, {x+1, y, s}, {x,y,s}
        ) + HC;
      }
      if(y+1 < SIZE_Y){
        model << HoppingAmplitude(
          1, {x,y+1, s}, {x,y, s}
        ) + HC;
      }
    }
  }
}
\end{lstlisting}
Here "+ HC" implies that both the hopping amplitude and its Hermitian conjugate is added to the model.
By allowing the user to specify the model using physical indices with arbitrary structure, it is relatively easy to specify virtually any Hamiltonian of interest given that the $a_{\mathbf{i}\mathbf{j}}$'s are known.
In fact, it is even possible to specify Hamiltonians that are time dependent or which depend on some yet undetermined parameters by passing so called callback functions as the first parameter to the hopping amplitudes.
The later can be particularly useful if for example some parameters are to be determined self-consistently or if they need to be calculated as some yet unknown overlap integrals.
For more information on this we refer to the documentation\footnote{With doxygen installed, the documentation can be built using {\it make documentation} and is then found at BuildFolder/doc/html/index.html, where BuildFolder is the build folder. The documentation for the latest release is also available at http://www.second-quantization.com. \label{Footnote:Documentation}}.

\begin{figure}
\includegraphics[width=250pt]{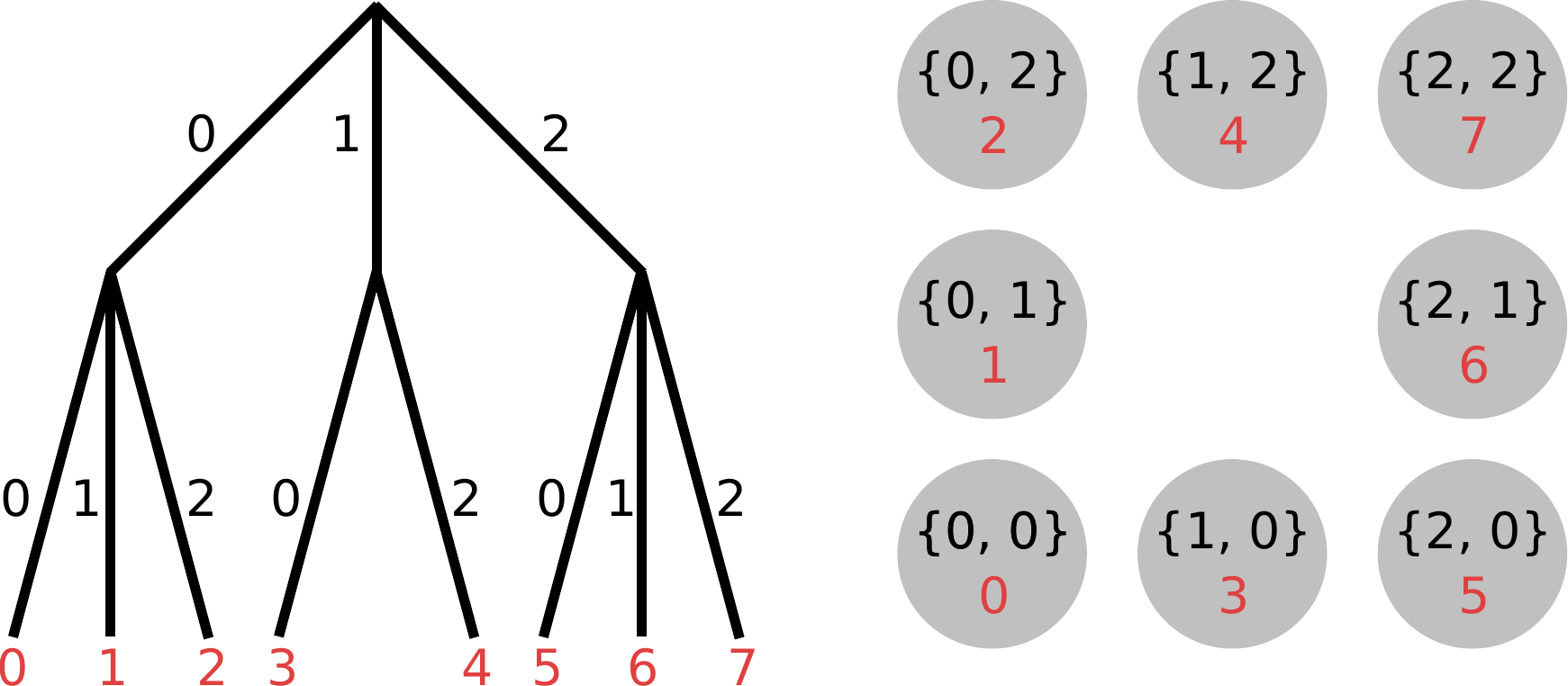}
\caption{As hopping amplitudes are added to the Model, a tree structure is being built. The hopping amplitude is passed down along the branches of the tree and stored on the leaf nodes, with the nth subindex of the second hopping amplitude index determining which branch to go down at the nth level. If hopping amplitudes consistent with the geometry on the right side of the figure are added to the model, the resulting tree structure is as shown on the left. When the model.construct() call is made, a linear Hilbert space basis is created by traversing the tree and enumerating the leaf nodes with increasing numbers starting from 0.}
\label{Fig:TreeStructure}
\end{figure}

Once all relevant hopping amplitudes are added to the model, the mapping between the physical indices and the Hilbert space indices are created using
\begin{lstlisting}[language=c++]
model.construct();
\end{lstlisting}
Taking a closer look at what happens behind the scenes of this call is useful for understanding why we can afford the added convenience of physical indices without incurring a significant performance penalty.
The hopping amplitudes are stored in a tree structure, see Fig.~\ref{Fig:TreeStructure}, which can be thought of as a sparse matrix format with each column stored on a leaf node.
The physical indices associated with each column is encoded in the tree structure itself, while the linear Hilbert space index is stored on the leaf node once the model.construct() call has been made.
In addition to acting as a storage for the hopping amplitudes, the tree structure therefore also provides a mapping between the physical and Hilbert space indices.
Note that since the maping only includes sites that are actually included in the model, this results in a minimal Hilbert space.

Since a physical index can be converted to a Hilbert space index by traversing the tree to the corresponding leaf node, the time complexity for a conversion from a physical to a Hilbert space index is $O(1)$ in the Hilbert space size.
The time complexity for a single reverse lookup is somewhat more complicated since it depends on the number of branches per level, but algorithms that need a reverse lookup can construct a lookup table in $O(N)$ time by iterating through the tree structure, where $N$ is the Hilbert space size.
Of particular interest to the method developer is the fact that it is possible to iterate through all the hopping amplitudes and extract their value in the Hilbert space basis in a time that is $O(M)$, where $M$ is the number of hopping amplitudes.
This is virtually always also linear in the Hilbert space size since the number of hopping amplitudes is proportional to the Hilbert space size for any model with local operators (which is the case for tight-binding, finite differences, finite elements, etc.).

Although we do not go into details about any of the preliminary support for many particle systems, we note that the tree structure described above describes the (bilinear) single particle part of the Hamiltonian.
Interaction terms are stored in a separate structure, but the mapping provided by the tree structure allows also for the operators in the interaction terms to be refered to using their physical and Hilbert space indices interchangably.

When relevant, the model object can also hold information such as the temperature, chemical potential, and statistics
\begin{lstlisting}[language=c++]
model.setTemperature(300);
model.setChemicalPotential(0);
model.setStatistics(Statistics::FermiDirac);
\end{lstlisting}
The model object is thus the general purpose container for model related information and can also contain other information not shown here.

\section{Solvers}
Different type of problems require different solution methods and these are generally implemented in solvers in TBTK.
Since the solver usually is where most computational time is spent, it is important that method developers have complete freedom over the choice of data structures and programing paradigm that are used internally to implement their algorithms.
Moreover, for general purpose solvers it is also important that they can work with minimal assumptions about the model.
The mapping from physical indices to Hilbert space indices provides the key to solving both of these problems.
Internally the solvers can request the hopping amplitudes from the model using the Hilbert space basis and set up the data structures that are best suited for the method.
In this way method developers can create new solvers without worrying about the details of specific physical models, while application developers can specify models without worrying about the method specific details of particular solvers.

TBTK comes packed with a number of different solvers that for example can perform diagonalization, Arnoldi iteration, and Chebyshev expansion of the Green's function \cite{PhysRevLett.105.167006}.
Using diagonalization as an example, a typical solver can be set up and executed as follows.
\begin{lstlisting}[language=c++]
Solver::Diagonalizer solver;
solver.setModel(model);
solver.run();
\end{lstlisting}
For other solvers the initialization may require more method specific parameters to be supplied.
However, the main idea is to provide an interface to the application developer that minimizes the amount of necessary method specific knowledge while simultaneously providing the possibility to configure the solver on demand.

Since the Hamiltonian can be converted to whatever format that is most suitable for the algorithm in $O(N)$ time and the execution time of almost every solution method is superlinear in $N$, the penalty for this conversion is negligible.
The basic recipe a method developer can apply when setting up the calculation inside the solver is as follows.
\begin{lstlisting}[language=c++]
//Iterate over the HoppingAmplitudes.
const HoppingAmplitudeSet
  &hoppingAmplitudeSet
    = model.getHoppingAmplitudeSet();
for(
  HoppingAmplitudeSet::ConstIterator
    iterator
      = hoppingAmplitudeSet.cbegin();
  iterator != hoppingAmplitudeSet.cend();
  ++iterator
){
  //Extract the amplitude and physical
  //indices from the HoppingAmplitude.
  complex<double> amplitude
    = (*iterator).getAmplitude();
  const Index &toIndex
    = (*iterator).getToIndex();
  const Index &fromIndex
    = (*iterator).getFromIndex();

  //Convert the physical indices to linear
  //indices.
  int row
    = hoppingAmplitudeSet.getBasisIndex(
      toIndex
    );
  int column
    = hoppingAmplitudeSet.getBasisIndex(
      fromIndex
    );

  //Add the matrix element to the
  //Hamiltonian on the format best suited
  //for the given algorithm.
  //...
}
\end{lstlisting}

\section{Extracting properties}
TBTK defines several properties such as eigenvalues, wave functions, density of states (DOS), (spin-polarized) local density of states (LDOS), etc.
However, different solvers can internaly use very different storage structures and it is desirable to limit the solvers responsibility to dealing with the general purpose problem formulated using Hilbert space indices.
For this reason TBTK provides property extractors that bridge the gap between the method specific details of the solvers and the higher abstraction layer presented to application developers.
Method developers are strongly advised to create similar property extractors in parallel with their solvers.

The property extractors provide a more intuitive interface to the application developer, allowing the application developer to extract properties from the solvers using physical indices.
Moreover, they aim to provide uniform interfaces for the solvers to the outside world.
Code that uses property extractors can therefore often work even if the solver is changed and makes it possible to try completely different solution method by simply changing a few lines of code related to the solver initialization.
We do, however, note that not every solver can calculate every property, and some solvers can calculate some specific details that are not available through other solvers at all.
Property extractors are therefore only approximately uniform, sometimes providing implementations for functions that simply print that the corresponding solver cannot be used to calculate the given property, while sometimes having additional functions not available in other property extractors.

A typical expression for seting up a property extractor is as follows.
\begin{lstlisting}[language=c++]
PropertyExtractor::Diagonalizer
  propertyExtractor(solver);
propertyExtractor.setEnergyWindow(
  LOWER_BOUND,
  UPPER_BOUND,
  RESOLUTION
);
\end{lstlisting}
We can then extract the density on each site, summing over spins:
\begin{lstlisting}[language=c++]
Property::Density density
  = propertyExtractor.calculateDensity(
    {{IDX_ALL, IDX_ALL, IDX_SUM_ALL}}
  );
\end{lstlisting}
Calculate the density of states (DOS):
\begin{lstlisting}[language=c++]
Property::DOS dos
  = propertyExtractor.calculateDOS();
\end{lstlisting}
Get the eigenvalues:
\begin{lstlisting}[language=c++]
Property::EigenValues eigenValues
  = propertyExtractor.getEigenValues();
\end{lstlisting}
Calculate the retarded Green's function for all spin combinations, with the annihilation operator on site (0, 0) and the creation operator on site (5, 5):
\begin{lstlisting}[language=c++]
Property::GreensFunction greensFunction
  = propertyExtractor.calculateGreensFunction(
    {{{0, 0, IDX_ALL}, {5, 5, IDX_ALL}}},
    Property::GreensFunction::Type::Retarded
  );
\end{lstlisting}
Extract the LDOS along y = SIZE\_Y/2, summing over spins:
\begin{lstlisting}[language=c++]
Property::LDOS ldos
  = propertyExtractor.calculateLDOS(
    {{IDX_ALL, SIZE_Y/2, IDX_SUM_ALL}}
  );
\end{lstlisting}
Calculate the magnetization on all sites:
\begin{lstlisting}[language=c++]
Property::Magnetization magnetization
  = propertyExtractor.calculateMagnetization(
    {{IDX_ALL, IDX_ALL, IDX_SPIN}}
  );
\end{lstlisting}
Get the spin-polarized LDOS on site (2, 4) and (3, 5):
\begin{lstlisting}[language=c++]
Property::SpinPolarizedLDOS
  spinPolarizedLDOS
    = propertyExtractor.calculateSpinPolarizedLDOS(
      {
        {2, 4, IDX_SPIN},
        {3, 5, IDX_SPIN},
      }
  );
\end{lstlisting}
Extract the wavefunction for all indices for state 1, 3, and 7:
\begin{lstlisting}[language=c++]
Property::WaveFunctions waveFunctions
  = propertyExtractor.calculateWaveFunctions(
    {{IDX_ALL, IDX_ALL, IDX_ALL}},
    {1, 3, 7}
  );
\end{lstlisting}

\section{Benchmark}
We perform a few benchmarks to quantify the scaling behavior and provide convincing evidence that the general purpose data structures introduced comes with negligible performance penalties.
In particular, the model specification and retrieval of the Hamiltonian on a sparse matrix format is considered since this is the layer that separates application developer code from method developer code.
These benchmarks therefore provides an upper bound on the overhead cost of the data structures.
The benchmark is done against kwant~\cite{Kwant} since it is able to achieve the same things, although using a less general formalsim where coordinates and other indices are treated on different footing.
See PerformanceTest.zip in the supplemental material for the actual code used.

\begin{figure}
\includegraphics[width=200pt]{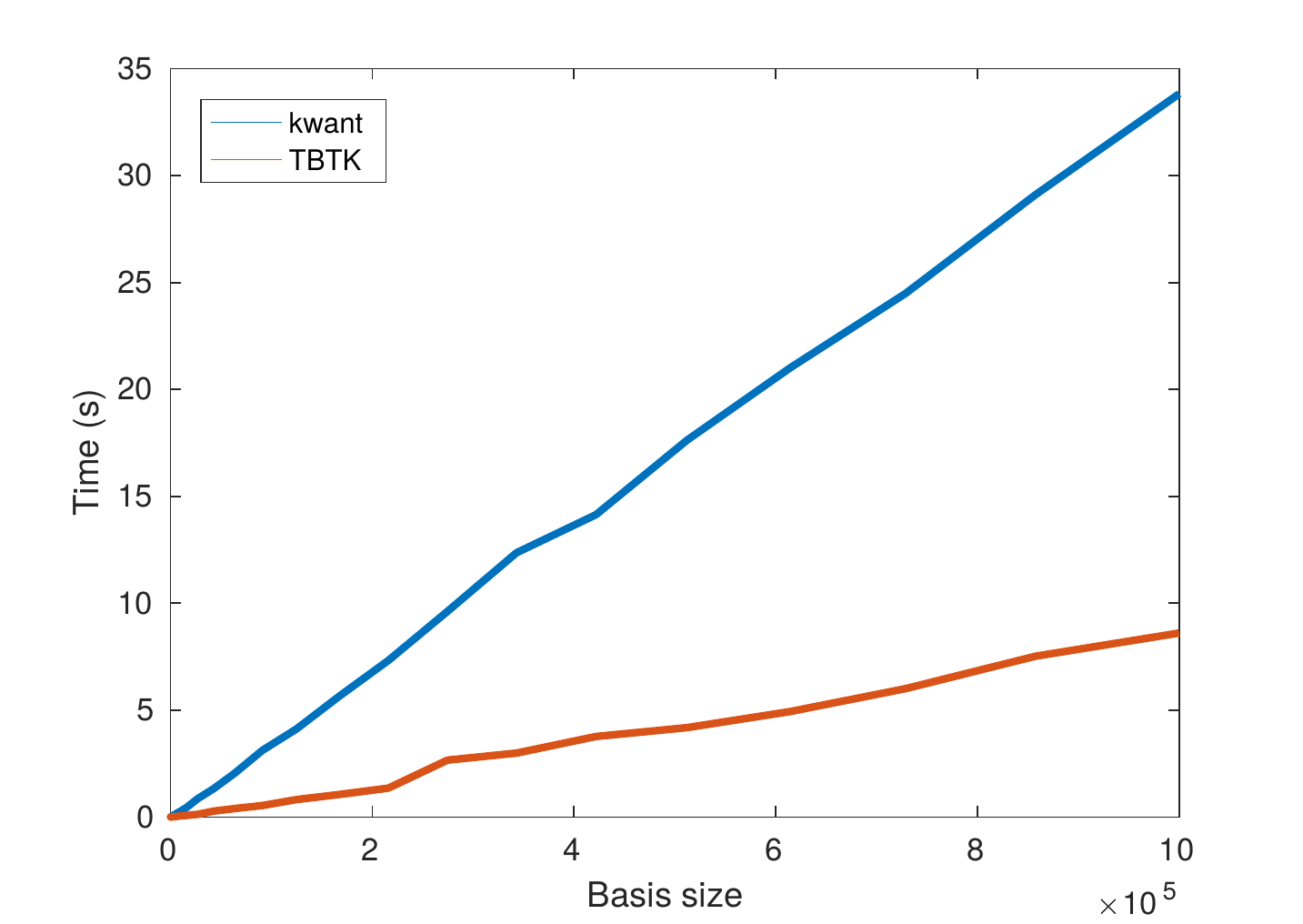}
\includegraphics[width=200pt]{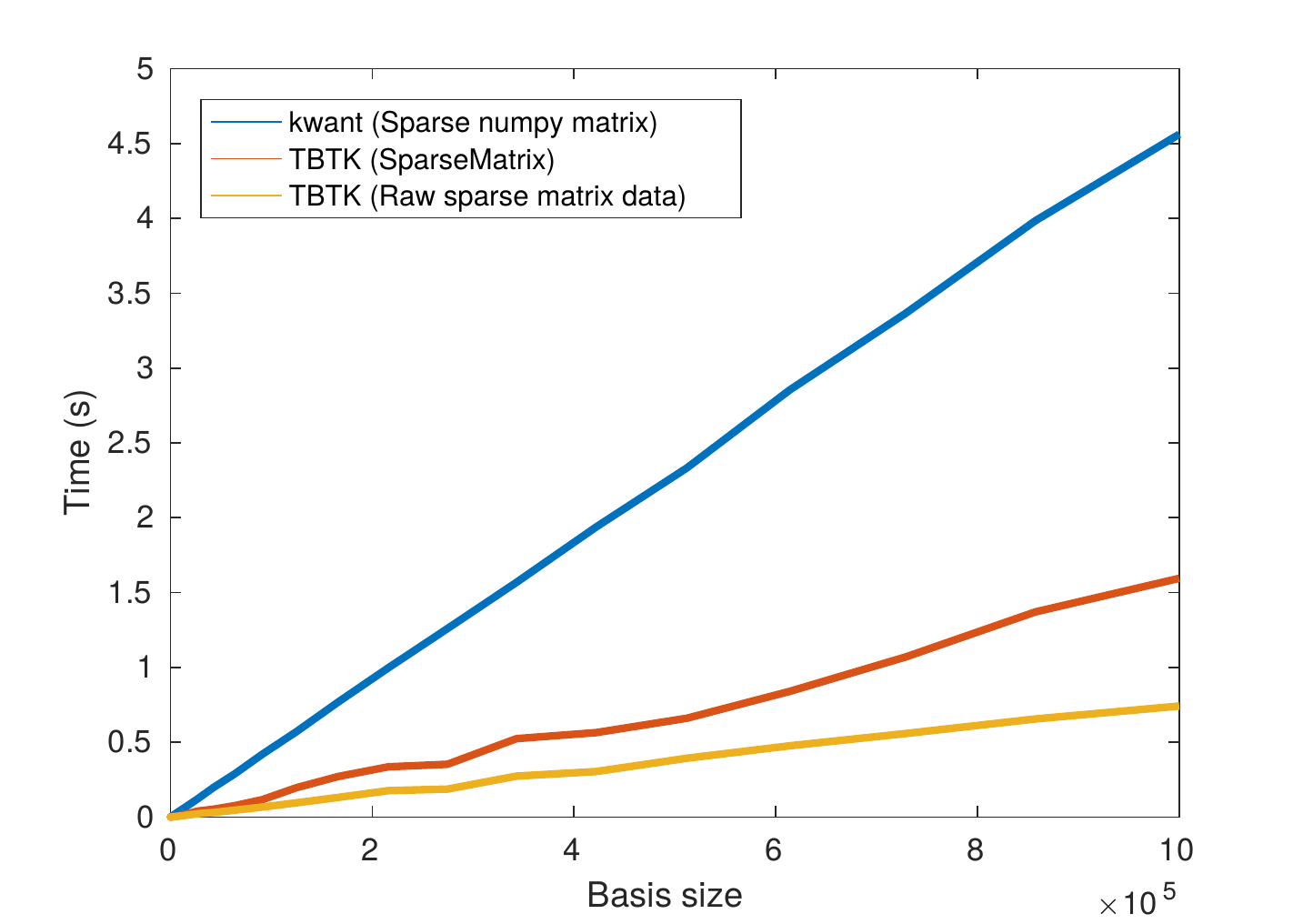}
\caption{(Left) Time spent setting up an $N\times N\times N$ cubic tight binding-model with onsite and nearest neighbor hopping terms, as a function of the Hilbert space basis size. TBTK is approximately four times faster ($3.93$ times faster at basis size $10^6$). (Right) Time spent extracting the same Hamiltonian, as a function of the Hilbert space basis size. Extracting the matrix as a SparseMatrix object in TBTK is about three times faster than extracting it as a sparse numpy array in kwant (2.86 times faster at basis size $10^6$). Method developers can squeeze out an additional factor of two in TBTK by accessing the raw sparse matrix data directly (6.15 times faster at basis size $10^6$).}
\label{Fig:Benchmark}
\end{figure}

For simplicty a cubic tight-binding model with nearest neighbor hopping is created and then converted to a sparse matrix on a fresh install of Ubuntu 18.04 running on a single core Intel(R) Xeon(R) CPU @ 2.30 GHz.
In the left panel of Fig.~\ref{Fig:Benchmark} the time to set up a model as a function of the Hilbert space size is shown.
Both TBTK and kwant display linear scaling, with TBTK outperforming kwant by a factor of almost 4.
In the right panel of Fig.~\ref{Fig:Benchmark} the time for extracting the Hamiltonian on a sparse matrix format is ploted.
Also here both TBTK and kwant scales linearly, with TBTK beating kwant by a factor of 3 to 6 depending on the specific sparse format chosen.
In either case, the overhead cost for setting up a model and requesting the data is not particularly large for either TBTK or kwant.
In fact, for problems with a basis size as large as $10^6$, the commulative time of about 10 seconds for TBTK and 38 seconds for kwant to specify and extract the model on a sparse matrix format is almost certainly going to be dwarfed by the time spent in any algorithm working on a system of that size.

\section{Impact and summary}
TBTK is a software development kit that enables rapid development of applications that calculates quantum mechanical properties.
It will aid the scientific community and industry in developing codes that enable large scale collaborations through a scalable approach that encourages reusability.
A more integrated community is essential for quantum technology to scale to an industrial level.
It also means that errors etc. can be detected more quickly and provides a means towards the transparency that scientific work is supposed to have.
This will allow the community to spend less time replicating numerical details that others already have worked out and to put more focus on the physical questions of interest.
To achieve this, TBTK provides a set of efficient general purpose data structures and build tools that draws from the latest best practices in software development.
The intention is to enable the development of an ecosystem of solvers and tools that can perform tight-binding, DFT, DMFT, quantum transport, and other types of calculations and to make it easy to integrate the different methods with each other.
In particular, TBTK aims to aid such development by providing data structures that allow developers to work at a higher level of abstraction, enabling them to put more focus on the physical ideas than on numerical details.
The interested reader is referred to the documentation\footnote{See footnote \ref{Footnote:Documentation}.} for more information.

\section{Acknowledgements}
Thanks to Anna Sinelnikova, Igor Di Marco, Mahdi Mashkoori, Oladunjoye Awoga, Andreas Theiler, and Tom McClintock for useful discussions and feedback on the code and/or manuscript.
Also thanks to Annica M. Black-Schaffer and Brian M. Andersen for fruitful collaborations that has helped shape the code.
The development has been made possible through financial support from the Swedish Research Council (Vetenskapsr\r{a}det, Grant No. 621-2014-3721), the G\"{o}ran Gustafsson Foundation, the Swedish Foundation for Strategic Research (SSF), the Knut and Alice Wallenberg Foundation through the Wallenberg Academy Fellows program, and the Independent Research Fund Denmark grant number DFF-6108-00096.





\clearpage

\begin{table}[]
\begin{tabular}{|l|l|l|}
\hline
N r & (executable) Code metadata & \\
N r & description & \\
\hline
C 1 & Current code version & v1.0.3\\
\hline
C 2 & Permanent link to & https://github.com/dafer45/TBTK/releases/tag/v1.0.3\\
~ & code / repository used & ~\\
~ & for this code version & ~\\
\hline
C 3 & Legal code license & Apache License 2.0\\
\hline
C 4 & Code version system used & git\\
\hline
C 5 & Software Code Language used & C++, python\\
\hline
C 6 & Compilation requirements, & BLAS, LAPACK, CMake, \it{(optional: ARPACK,}\\
~ & Operating environments & \it{FFTW3, OpenCV, cURL, SuperLU (v5.2.1),}\\
~ & \& dependencies & \it{wxWidgets, CUDA, HDF5, OpenBLAS, OpenMP,}\\
~ & ~ & \it{Google Test)}\\
~ & ~ & Linux, OS X, Unix-like\\
\hline
C 7 & If available Link to developer & www.second-quantization.com\\
~ & documentation / manual & ~\\
\hline
C 8 & Support email for questions & kristofer.bjornson@gmail.com\\
\hline
\end{tabular}
\end{table}

\end{document}